\def\r#1{\mbox{\bf r}_{#1}}
\def\rp#1{\mbox{\bf r}_{#1}^{\prime}}
\def\d{\mbox{d}}
\documentstyle[aps,multicol,pre]{revtex}

\begin{document}
\draft
\title{\bf Pair Wave Functions in a Bose Liquid}
\author{A. Yu. Cherny~\cite{email}}
\address{Bogoliubov Laboratory of Theoretical Physics,
Joint Institute for Nuclear Research, 141980, Dubna, Moscow region,
Russia}
\date{February 8, 2000}

\maketitle

\begin{abstract}
Pair wave functions (PWF) which are eigenfunctions of the reduced density
$2$-matrix are considered for a homogeneous Bose liquid.  With the Bogoliubov
principle of the correlation weakening it is proved that the distribution of
the ``dissociated" pair states over momenta is exactly the product of the
single-particle distribution functions. Thus, the ``dissociated" pair states
are naturally classified as condensate-condensate, condensate-supracondensate
and supracondensate-supracondensate ones provided the Bose-Einstein
condensate exists.  The condensate-condensate as well as
condensate-supracondensate PWF are expressed in terms of the averages of
products of the creation and destruction Bose operators.  This leads to the
simple interpretation of the anomalous averages as the ``scattering parts" of
the condensate-condensate and condensate-supracondensate PWF.  It is shown
that in contrast to the Fermi liquid, the appearance of the anomalous
averages for the Bose liquid does not necessarily mean that there exist bound
states of pairs of particles.  The PWF in the Hartree-Fock-Bogoliubov (HFB)
approach are found.  Given the density of the condensate is not zero, there
are no bound pair states in the HFB scheme.  The expansion of the pair
correlation function in the set of PWF is very useful in order to take into
account both short-range and long-range spatial correlations.  Applications
(possible and already realized) of the formalism developed are discussed.
\end{abstract}

\begin{multicols}{2}

\section{Introduction}
\label{1}

It is well-known that for the homogeneous Fermi liquid the appearance of
anomalous averages $\langle{\hat\psi}^{\dagger}(x_{1})
{\hat\psi}^{\dagger}(x_{2})\rangle$ implies that there exists a macroscopic
number of bound pairs of particles with zero momentum (here
${\hat\psi}^{\dagger}(x)$ and ${\hat\psi}(x)$ are the field operators). In
the case of the pair condensate like this, the properties of the system
change radically. In particular, this leads to the superfluidity of the Fermi
liquid (superconductivity in an electron system, superfluidity in $^{3}$He).

For a homogeneous Bose liquid, the superfluidity is usually associated with
the macroscopic number of particles themselves in the zero-momentum state
(the Bose-Einstein condensate). In the Bogoliubov theory for a weakly
imperfect Bose gas, repulsive interaction between particles depletes the
condensate, while makes it thermodynamically stable.  In this case the
anomalous averages $\langle{\hat\vartheta}^{\dagger}(\r{1})
{\hat\vartheta}^{\dagger}(\r{2})\rangle$ play a crucial role also, where
${\hat\vartheta}^{\dagger}({\bf r})$ is the field operator creating a
supracondensate particle at a point ${\bf r}$, see Eq. (\ref{thetakrr}).

Parallels between the Bogoliubov transformations for boson~\cite{bog47} and
fermion~\cite{bog58} systems gave birth to pair theories being special cases
of the HFB theory (see, e.g. Ref.~\cite{kobe} and references therein). In the
framework of the HFB approach, the gap in the single-particle excitation
energy appears~\cite{hohmartin} that may be interpreted, by analogy with the
theory of superconductivity, as an evidence for the bound states of pairs of
particles.  Below we show that in the presence of a single-particle
condensate, there are no bound pair states of bosons in the HFB theory. Thus,
it has been verified in terms of PWF that the gap is unphysical, which has
been stressed in many papers~\cite{hohmartin,hugpines}.

The concept of wave functions for a group of $m$ particles in medium can be
rigorously introduced with eigenfunctions of the $m$-th-order reduced density
matrix, or the $m$-matrix.  Indeed, the system of $m$ particles is a
subsystem of that of $N$ particles. So, its state is not pure even in the
situation when the system as a whole has a wave function. In general a
subsystem is specified by the density matrix~(see, e.g. Ref.~\cite{Landau}).
The $m$-matrix is of use when we have a noncoherent superposition of the
$m$-particle wave functions.  L\"owdin and Shull used wave functions like
these for one and two particles to describe electron states in atoms, and
called them the ``natural orbitals" and ``natural geminals", respectively
(see Ref.~\cite{coleman}). The concept of PWF for fermions has been used by
Shafroth and his coworkers in papers on the theory of superconductivity (for
a review, see Ref.~\cite{blatt}). In what follows, we shall use the term PWF
for the wave functions of particle pairs, using the notation of Bogoliubov
who made the most clear presentation of the concept of PWF for
fermions~\cite{bogquasi}.

This paper concerns PWF for the Bose liquid below the temperature of the
condensation, $T_{c}$. Analysis is carried out in general form as far as it
is possible, with a special emphasis on the bound pair states and that part
of PWF which is responsible for the correlations of particles at short
distances. Consideration of these correlations is of special interest when
investigating concrete boson systems for which the potential of two-particle
interaction $\Phi(r)$ is, as a rule, strongly singular, or, in other words,
of the hard-core type (i.e. potential goes to infinity like $1/r^{m}$ ($m>3$)
as $r\rightarrow 0$, and, consequently, there is no Fourier transform of it).
For example, the well-known Lennard-Jones potential is of this form. In this
case standard approximations based on the weak-coupling pertrubation theory
lead to divergencies. For example, the statistical average of the
interaction energy per particle
\begin{equation}
\varepsilon_{int}=\frac{1}{N}\Bigl\langle\frac{1}{2}\sum_{i\not=j} \Phi(|{\bf
r}_{i}-{\bf r}_{j}|)\Bigr\rangle=\frac{n}{2}\int\d^{3}r\,\Phi(r)g(r)
\label{uaver}
\end{equation}
(here $g(r)$ is the pair distribution function, $n$ is the density of the
particles) is infinite in the simplest Hartree approximation corresponding to
$g(r)=1$.  A reason for the divergency is obvious as the short-range spatial
correlations of two particles are not properly taken into consideration.  In
order to escape the mentioned trouble a real potential is usually replaced by
various effective potentials for which the Fourier transforms exist (for the
Bose case see the $t$-matrix approach~\cite{brueck}, the pseudopotential with
positive scattering length, or the hard-sphere model,~\cite{lee} and the
method by Beliaev~\cite{bel}). This replacement implies that the effective
potential contains all ``information" on the short-range correlations of
particles. A range of validity of this replacement for the many-boson systems
is discussed in the textbook~\cite{popov} and, for an inhomogeneous gas, in
the recent paper~\cite{burnett} (see the discussion at the end of
Sec.~\ref{3a}).  In the lowest order in density $n$ of the Bose gas such
effective-interaction procedures are reduced to the Bogoliubov model with the
effective potential [for example, in the pseudopotential method the Fourier
transform of the ``bare" potential is replaced by $\Phi(k)=4\pi a/m$, where
$a$ is the scattering length obtained from the two-body Schr\"odinger
equation].  However, this approach  is found to be thermodynamically
inconsistent~\cite{CSPRE}, which can manifest itself in various unphysical
results. For example, one can obtain no depletion of the Bose condensate
within the pseudopotential approach, if one tries to apply a self-consistent
treatment~\cite{huang}. Besides, we are not able to obtain the correct value
of the interaction energy (\ref{uaver}) by direct calculations using
$g(r)$~\cite{CSEPJB}. Thus, it looks interesting and promising to realize an
alternative variational scheme that is self-consistent from the very
beginning, as the short-range correlations are properly taken into account
for all the terms in the expansion of $g(r)$ (\ref{gr}) in the set of PWF.
Thus, PWF are ``regular channels" which can be used to cancel the
divergencies in a self-consistent manner (see discussion in Secs.~\ref{3a}
and \ref{4}). The first results derived in Refs.~\cite{CSPRE,CSJPS,CSPRE2}
demonstrate that the formalism of the PWF can actually be employed as a basis
for the self-consistent variational treatment for a Bose gas with the
strongly singular interaction potential.

General properties of the eigenfunctions and eigenvalues of the correlation
function (\ref{f2}) (differing from the $2$-matrix only by a norm) are
interesting in itself when an eigenvalue becomes a macroscopic quantity
(off-diagonal long-range order~\cite{yang}). With the Bogoliubov principle of
correlation weakening we shall show that eigenvalues of the correlation
function (\ref{f2}) which belong to continuous spectrum (``dissociated" pair
states) are expressed as the product of the single-particle momentum
distributions. Thus, the ``dissociated" pair states are naturally classified
as condensate-condensate, condensate-supracondensate and
supracondensate-supracondensate ones provided the Bose-Einstein condensate
exists. The condensate-condensate as well as condensate-supracondensate PWF
are expressed in terms of the averages of products of the creation and
destruction Bose operators.  This leads to the simple interpretation of the
anomalous averages as the ``scattering parts" of the condensate-condensate
and condensate-supracondensate PWF (see Sec.~\ref{3}).

In the next section, the PWF are introduced for boson systems by analogy with
Bogoliubov's paper~\cite{bogquasi}, and the continuous spectrum of the
correlation function (\ref{f2}) is calculated.  In Sec. \ref{3} we consider
the general structure of the $2$-matrix below $T_{c}$, the PWF in an explicit
form and their physical interpretation. In Sec.~\ref{3a} the PWF are
evaluated in HFB approach as well as in the Bogoliubov model, and some
applications of the developed formalism are considered. Main results are
summarized in Sec. \ref{4}.

\section{The concept of particle pair states for bosons}
\label{2}

Let us consider a homogeneous system of $N$ bosons, and, for simplicity, let
the spin of the particles be equal to zero. The Hamiltonian is assumed to be
invariant with respect to translations and the transformation of particle
momentum ${\bf p}_{i}\rightarrow -{\bf p}_{i}$.  The latter implies that the
Hamiltonian does not change when the canonical transformation ${\hat a}_{{\bf
k}}\to {\hat a}_{-{\bf k}},\ {\hat a}^{\dagger}_{{\bf k}}\to {\hat
a}^{\dagger}_{-{\bf k}}$ is performed. In particular, this condition is
satisfied for pairwise interactions with a potential $\Phi({\bf r})=
\Phi(r)$. A state of the system is determined by a density matrix
\begin{eqnarray}
\rho(\rp{1},\ldots,\rp{N};\r{1},\ldots,\r{N})=\sum_{n}
&&w_{n}\Psi_{n}^{*}(\r{1},\ldots,\r{N})
\nonumber \\
&&\times\Psi_{n}(\rp{1},\ldots,\rp{N}),
\label{rho}
\end{eqnarray}
where $\sum_{n}w_{n}=1$, $w_{n}\geq 0$, and $\Psi_{n}$ are orthonormal system
of functions being symmetric with respect to any permutations of particles.

The $2$-matrix is defined as
\begin{eqnarray}
\rho_{2}&&(\rp{1},\rp{2};\r{1},\r{2})=\int_{V}\d^{3}r_{3}\cdots \d^{3}r_{N}
\nonumber \\
&&\times\rho(\rp{1},\rp{2},\r{3},\ldots,\r{N};\r{1},\r{2},\r{3},\ldots,
\r{N}),
\label{rho2def}
\end{eqnarray}
and can be expressed in terms of the field operators ${\hat\psi}(\r{})$ and
${\hat\psi}^{\dagger}(\r{})$ (see, e.g. Ref.~\cite{boglec})
\begin{equation}
\rho_{2}(\rp{1},\rp{2};\r{1},\r{2})=\frac{1}{N(N-1)}
\langle{\hat\psi}^{\dagger}(\r{1}){\hat\psi}^{\dagger}(\r{2})
{\hat\psi}(\rp{2}){\hat\psi}(\rp1{})\rangle
\label{rho2vtor}
\end{equation}
Here $\langle\cdots\rangle=\mbox{Tr}\,(\cdots\hat\rho)$ stands for the
statistical average over the state $\hat\rho$. The density matrix
$\rho_{2}$ is normalized, namely
$$
\mbox{Tr}\,\rho_{2}=\int_{V}\d^{3}r_{1}\d^{3}r_{2}\;
\rho_{2}(\r{1},\r{2};\r{1},\r{2})=1,
$$
therefore, any matrix element (\ref{rho2vtor}) has asymptotic
behaviour as $1/V^{2}$, in the thermodynamic limit $V\rightarrow\infty$,
$N/V= n = \mbox{const}$.  Therefore, it is more convenient to
work with the pair correlation function
\begin{equation}
F_{2}(\r{1},\r{2};\rp{1},\rp{2})=
\langle{\hat\psi}^{\dagger}(\r{1}){\hat\psi}^{\dagger}(\r{2})
{\hat\psi}(\rp{2}){\hat\psi}(\rp1{})\rangle.
\label{f2}
\end{equation}

The boundary conditions for $F_2$~\cite{Note3a} follow from the
principle of the correlation weakening at macroscopical
separations~\cite{bogquasi}:
\begin{eqnarray}
\langle {\hat\psi}^{\dagger}({\bf r}_1) {\hat\psi}^{\dagger}({\bf r}_2)
&&{\hat\psi}({\bf r}_2^{\prime}){\hat\psi} ({\bf r}_1^{\prime})\rangle\to
\nonumber \\
&&\langle{\hat\psi}^{\dagger}({\bf r}_1) {\hat\psi}^{\dagger}({\bf r}_2)\rangle
\langle{\hat\psi} ({\bf r}_2^{\prime}){\hat\psi} ({\bf r}_1^{\prime})\rangle
\label{corr1}
\end{eqnarray}
when
\begin{equation}
{\bf r}_1 - {\bf r}_2 = \mbox{const},\
{\bf r}_1^{\prime} - {\bf r}_2^{\prime} = \mbox{const},\
|{\bf r}_1^{\prime} - {\bf r}_1| \to \infty ;
\label{limit1}
\end{equation}
\begin{eqnarray}
\langle {\hat\psi}^{\dagger}({\bf r}_1) {\hat\psi}^{\dagger}({\bf r}_2)
&&{\hat\psi}({\bf r}_2^{\prime}){\hat\psi} ({\bf r}_1^{\prime})\rangle\to
\nonumber \\
&&\langle {\hat\psi}^{\dagger}({\bf r}_1) {\hat\psi}({\bf r}_1^{\prime})\rangle
\langle{\hat\psi}^{\dagger} ({\bf r}_2){\hat\psi} ({\bf r}_2^{\prime})\rangle
\label{corr2}
\end{eqnarray}
when
\begin{equation}
{\bf r}_1 - {\bf r}_1^{\prime} = \mbox{const},\
       {\bf r}_2 - {\bf r}_2^{\prime} = \mbox{const},\
|{\bf r}_1 - {\bf r}_2| \to \infty.
\label{limit2}
\end{equation}
It should be stressed that limits (\ref{corr1}) and (\ref{corr2}) are valid
either the Bose-Einstein condensation takes place or not. In the first case
the anomalous averages in the expression (\ref{corr1}) are not
zero~\cite{note7a}.

As the expression (\ref{f2}) is a Hermitian kernel, we can expand it
in the orthonormal set of its eigenprojectors (in the Hilbert space):
\begin{equation}
F_{2}(\r{1},\r{2};\rp{1},\rp{2})=
\sum_{\nu}N_{\nu}\Psi_{\nu}^{*}(\r{1},\r{2})\Psi_{\nu}(\rp{1},\rp{2})
\label{f2psinu}
\end{equation}
where
\begin{equation}
\int_{V}\d^{3}r_{1}\d^{3}r_{2}\;{|\Psi_{\nu}
(\r{1},\r{2})|}^{2}=1.
\label{psinorm}
\end{equation}
Eigenfunctions $\Psi_{\nu}(\r{1},\r{2})$, which at the same time are
eigenfunctions of $2$-matrix (\ref{rho2vtor}), are called the wave functions
of pairs of particles, or PWF. From (\ref{f2}), (\ref{f2psinu}) and
(\ref{psinorm}) we get
$$
\int_{V}\d^{3}r_{1}\d^{3}r_{2}\;F_{2}(\r{1},\r{2};\r{1},\r{2})=
\langle {\widehat N}^{2}-{\widehat N}\rangle
$$
$$
=N(N-1)=\sum_{\nu}N_{\nu},
$$
i.e. the sum of all $N_{\nu}$ is the total number of pairs.  Therefore, the
non-negative quantities $N_{\nu}$ can be interpreted as the mean number of
the pairs in the state $\nu$, any pair being doubly taken. The ratio
$w_{\nu}=N_{\nu}/[N(N-1)]$ is the probability of observing a particle pair in
the pure state with the wave function $\Psi_{\nu}({\bf r}_1, {\bf r}_2)$.
Here, as one might expect, $\sum_{\nu} w_{\nu}=1$.

From the definition (\ref{f2}) it follows that
$$
F_{2}(\r{1},\r{2};\rp{1},\rp{2})=
F_{2}(\r{2},\r{1};\rp{1},\rp{2})=
F_{2}(\r{1},\r{2};\rp{2},\rp{1})
$$
hence, $\Psi_{\nu}(\r{1},\r{2})=\Psi_{\nu}(\r{2},\r{1})$, i.e. the
PWF for bosons are symmetric with respect to permutations.

For equilibrium states, the momentum of centre of mass of a pair of particles
${\bf q}$ is a quantum number corresponding to PWF, so the index $\nu$ can be
represented as $\nu=(\omega,{\bf q})$, where $\omega$ stands for other
indices. PWF can be written in the following form
\begin{equation}
\Psi_{\nu}(\r{1},\r{2})=\psi_{\omega,{\bf q}}(\r{1}-\r{2})
\frac{\exp [i {\bf q}(\r{1}+\r{2})/2]}{\sqrt{V}}.
\label{psinurazd}
\end{equation}
Let us demonstrate this. Consider an equilibrium state described by the Gibbs
canonical ensemble. As the Hamiltonian is invariant under translations, the
total momentum ${\bf P}$ is conserved. So, we have
\begin{equation}
\Psi_{n}(\r{1}+{\bf a},\ldots,\r{N}+{\bf a})=
\Psi_{n}(\r{1},\ldots,\r{N})\exp[i{\bf P}{\bf a}],
\label{psip1}
\end{equation}
Without loss of generality we put ${\bf P}=0$, and introduce new variables
\begin{equation}
{\bf R}=(\r{1}+\r{2})/2,\quad {\bf r}=\r{1}-\r{2},
\label{Rr}
\end{equation}
and similarly for $\rp{1}$ and $\rp{2}$. Rewrite (\ref{rho}) with the help of
(\ref{psip1}) setting ${\bf a}={\bf R}$ and ${\bf a}={\bf R}^{\prime}$,
respectively. Substituting rewritten Eq.  (\ref{rho}) into (\ref{rho2def})
and performing the Fourier transformations with respect to $\rp{3}-{\bf
R}^{\prime},\ldots,\rp{N}-{\bf R}^{\prime},\r{3}-{\bf R},\ldots,\r{N}-{\bf
R}$ we obtain:
\begin{equation}
\rho_{2}(\rp{1},\rp{2},\r{1},\r{2})=\frac{1}{V}\sum_{{\bf q}}
e^{i{\bf q}({\bf R}^{\prime}-{\bf R})}
\rho_{{\bf q}}(\rp{},\r{}),
\label{rhoq}
\end{equation}
where $\rho_{{\bf q}}$ is a non-negative Hermitian kernel. Expanding it in
the set of its eigenprojectors $\psi_{\omega,{\bf q}}({\bf
r}')\psi^{*}_{\omega,{\bf q}}(\r{})$ we arrive at Eqs.  (\ref{f2psinu}) and
(\ref{psinurazd}). Analogous proof can be given for the Gibbs grand canonical
ensemble.

Then the expression (\ref{f2psinu}) can be written in the form
\begin{eqnarray}
F_{2}(\r{1},\r{2};\rp{1},\rp{2})&=&\sum_{\omega,{\bf q}}
\frac{N_{\omega,{\bf q}}}{V}
\psi^{*}_{\omega,{\bf q}}(\r{1}-\r{2})\psi_{\omega,{\bf q}}(\rp{1}-\rp{2})
\nonumber \\
&&\times\exp\left[i\frac{{\bf q}}{2}(\rp{1}+\rp{2}-\r{1}-\r{2})\right].
\label{f2omq}
\end{eqnarray}
For $\psi_{\omega,{\bf q}}(\r{})$, which can be interpreted as the wave
function of a pair of particles in the centre-of-mass system, from
(\ref{psinorm}) and (\ref{psinurazd}) we get the following normalization
condition:
\begin{equation}
\int_{V}\d^{3}r\;|\psi_{\omega,{\bf q}}(\r{})|^{2}=1.
\label{psiomqn}
\end{equation}
As the wave function $\psi_{\omega,{\bf q}}(\r{})$ is an EF of the
kernel $\rho_{{\bf q}}(\rp{},\r{})$ (see Eq. (\ref{rhoq})), it can
belong either to a discrete or a continuous spectrum, so we have two
possible situations~\cite{note3}:\\
(1) $\psi_{\omega,{\bf q}}(\r{})\rightarrow 0$ at $r\rightarrow
\infty$ (bound state of a pair of particles).\\
(2) $\psi_{\omega,{\bf q}}(\r{})\rightarrow \sqrt{2}\cos ({\bf p}
{\bf r})$
at $r\rightarrow \infty$ (``dissociated" states of a pair of particles
corresponding to scattering with relative momentum ${\bf p}$, the factor
$\sqrt{2}$ stands for the appropriate normalization when $p\not=0$).

In the first case, $\omega=i$ is a discrete index numbering bound pair
states.  We denote $\psi_{\omega,{\bf q}}(\r{})=\varphi_{{\bf q}, i} (\r{})$
and keep the normalization (\ref{psiomqn}):
\begin{equation}
\int_{V}\d^{3}r\;|\varphi_{{\bf q},i}(\r{})|^{2}=1.
\label{svnorm}
\end{equation}
In the second case~\cite{note4}, $\omega={\bf p}$ and
$$
\psi_{\omega,{\bf q}}(\r{})=\frac{\varphi_{{\bf p},{\bf q}}
(\r{})}{\sqrt{V}}.
$$
Due to Eq. (\ref{psiomqn}) we obtain the normalization
\begin{equation}
\frac{1}{V}\int_{V}\d^{3}r\;|\varphi_{{\bf p},{\bf q}}
(\r{})|^{2}=1.
\label{freenorm}
\end{equation}
With the variables (\ref{Rr}) the expression (\ref{f2omq}) can be rewritten
in the following form:
\end{multicols}\vspace*{-3mm}\noindent\rule[2mm]{87.36mm}{.1mm}%
\rule[2mm]{.1mm}{2mm}\vspace*{-2mm}
\begin{eqnarray}
F_{2}(\r{1},\r{2};\rp{1},\rp{2})=F_{2}(\r{},{\bf R};
\rp{},{\bf R}^{\prime})
&=&\sum_{{\bf q},i}\frac{N_{q,i}}{V}
\varphi^{*}_{{\bf q},i}(\r{})\varphi_{{\bf q},i}(\rp{})
\exp[i{{\bf q}}({\bf R}^{\prime }-{\bf R})]
\nonumber \\
&&+\sum_{{\bf p},{\bf q}}\frac{N_{{\bf p},{\bf q}}}{V^{2}}
\varphi^{*}_{{\bf p},{\bf q}}(\r{})\varphi_{{\bf p},{\bf q}}
(\rp{})
\exp[i{{\bf q}}({\bf R}^{\prime }-{\bf R})].
\label{f2quasi}
\end{eqnarray}
In the limit $V\rightarrow \infty$ we should replace the quasidiscrete
momentum sums by the corresponding integrals:
\begin{eqnarray}
F_{2}(\r{},{\bf R};\rp{},{\bf R}^{\prime})
&=&\sum_{i}\int \d^{3}q\;w_{i}(q)
\varphi^{*}_{{\bf q},i}(\r{})\varphi_{{\bf q},i}(\rp{})
\exp[i{{\bf q}}({\bf R}^{\prime }-{\bf R})]
\nonumber \\
&&+\int \d^{3}p\,\d^{3}q\;w({\bf p},{\bf q})
\varphi^{*}_{{\bf p},{\bf q}}(\r{})\varphi_{{\bf p},{\bf q}}
(\rp{})
\exp[i{{\bf q}}({\bf R}^{\prime }-{\bf R})].
\label{f2asymp}
\end{eqnarray}
\hfill\vspace*{-2mm}\rule[0mm]{.1mm}{2mm}%
\rule[2mm]{87.36mm}{.1mm}\vspace*{-3mm}\begin{multicols}{2}
Thus, one can see from Eqs. (\ref{f2quasi}), (\ref{f2asymp}) that
$Vw_{i}(q)\d^{3}q$ is the number of bound pairs in state $i$ with momentum
${\bf q}$ in an infinitesimally small momentum volume $\d^{3}q$, and
$V^{2}w({\bf p},{\bf q})\d^{3}p\,\d^{3}q$ is the number of ``dissociated", or
scattered, pairs with relative momentum ${\bf p}$ and centre-of-mass momentum
${\bf q}$ in infinitesimally small momentum volumes $\d^{3}p$ and $\d^{3}q$.

In the centre-of-mass system the replacement ${\bf p} \to - {\bf p}$
corresponds to the permutation of particles. So, the following symmetric
relations take place:
\begin{equation}
w({\bf p},{\bf q})=w(-{\bf p},{\bf q}),
\label{wpqsimm}
\end{equation}
\begin{equation}
\varphi_{{\bf p},{\bf q}}({\bf r})=\varphi_{{\bf p},{\bf q}}(-{\bf r})
=\varphi_{-{\bf p},{\bf q}}({\bf r}).
\label{phisymm}
\end{equation}

Let us now prove, with the principle of the correlation weakening, that the
distribution of the particle pairs over the ``scattering" states $w({\bf
p},{\bf q})$ is expressed in terms of the single-particle momentum
distribution $w_{sing}({\bf k}) = n_{0} \delta({\bf k}) + n(k)/(2\pi)^{3}$,
where $\delta({\bf k})$ stands for the $3$-dimensional $\delta$-function,
$n(k)=\langle {\hat a}_{\bf k}^{\dagger} {\hat a}_{\bf k}\rangle$ is the
single-particle occupation number.  Suppose that the temperature is larger
than the temperature of the Bose-Einstein condensation, so, $n_{0}=0$ and
$w_{sing}(k)$ is a regular function, i.e. it does not contain
$\delta$-function after the thermodynamic limit.  On the one hand, in the
limiting situation of (\ref{limit2}) we have the relation (\ref{corr2}),
which can be written as
\end{multicols}\vspace*{-3mm}\noindent\rule[2mm]{87.36mm}{.1mm}%
\rule[2mm]{.1mm}{2mm}\vspace*{-2mm}
\begin{eqnarray}
F_{2}(\r{1},\r{2};\rp{1},\rp{2})
&\to&\int\frac{\d^3 p_1}{(2\pi)^3} n(p_1) \exp[i{\bf p}_1
({\bf r}_1^{\prime}-{\bf r}_1)]
\int\frac{\d^3 p_2}{(2\pi)^3} n(p_2) \exp[i{\bf p}_2
({\bf r}_2^{\prime}-{\bf r}_2)] \nonumber \\
&=&\int \d^3q\,\d^3p \frac{n({\bf q}/2+{\bf p})\,
         n({\bf q}/2-{\bf p})}{(2\pi)^6}
\exp[i{\bf p}({\bf r}^{\prime}-{\bf r})]
\exp[i{\bf q}({\bf R}^{\prime}-{\bf R})],
\label{F1F1}
\end{eqnarray}
where, passing to the last equality, we introduced the new variables ${\bf
q}={\bf p}_1+{\bf p}_2$ and ${\bf p}=({\bf p}_1- {\bf p}_2)/2$ and used
notations (\ref{Rr}). On the other hand, when (\ref{limit2}) is true, we have
$$
r=|{\bf r}_2-{\bf r}_1| \to \infty,\
   r^{\prime}=|{\bf r}_2^{\prime}-{\bf r}_1^{\prime}|\to \infty,
|{\bf r}+{\bf r}^{\prime}| \to \infty,\ {\bf R}^{\prime}-{\bf R}
=\mbox{const},\ {\bf r}^{\prime}-{\bf r}=\mbox{const}.
$$
\begin{multicols}{2}
Hence, in this limit we obtain
$$
\varphi_{{\bf q},i}(\r{})\rightarrow 0,\quad
\varphi_{{\bf p},{\bf q}}(\r{})\rightarrow \sqrt{2}\cos ({\bf p}{\bf r}).
$$
Therefore, it follows from (\ref{f2asymp}) that in the limiting case
$(\ref{limit2})$ we have
\begin{eqnarray}
F_{2}(\r{1},\r{2};\rp{1},\rp{2}) \to\int
&&\d^3 q\,\d^3 p\; w({\bf p},{\bf q})
2\cos({\bf p}{\bf r})\cos({\bf p}{\bf r}')
\nonumber \\
&&\times\exp[i{\bf q}({\bf R}^{\prime}-{\bf R})].
\label{F2lim}
\end{eqnarray}
Further, the Riemann's theorem~\cite{Note6} used while integrating over ${\bf
p}$ and relation (\ref{wpqsimm}) allow us to rewrite (\ref{F2lim}) as
\begin{eqnarray}
F_{2}(\r{1},\r{2};\rp{1},\rp{2}) \to \int
&&\d^3q\,\d^3p\;w({\bf p},{\bf q})
\exp\left[i{\bf p}({\bf r}^{\prime}-{\bf r})\right]
\nonumber \\
&&\times\exp\left[i{\bf q}({\bf R}^{\prime}-{\bf R})\right].
\label{F2lim2a}
\end{eqnarray}
The right-hand side of Eq. (\ref{F1F1}) is equal to that of (\ref{F2lim2a})
at all the values of space variables $\widetilde{\bf r}={\bf r}'-{\bf r}$ and
$\widetilde{\bf R}={\bf R}'-{\bf R}$.  Hence, we derive the following
equality:
\begin{equation}
w({\bf p},{\bf q})=\frac{n({\bf q}/2+{\bf p})
n({\bf q}/2-{\bf p})}{(2\pi)^6}.
\label{wpqsms}
\end{equation}

Below the temperature of the Bose-Einstein condensation a macroscopic number
of particles $N_{0}$ occupies the zero-momentum state:  $n_{0}=\langle {\hat
a}_{0}^{\dagger}{\hat a}_{0}\rangle/V =N_{0}/V =\mbox{const}$, when
$V\to\infty$, and we have in the thermodynamic limit
$$
\langle {\hat\psi}^{\dagger}({\bf r}_1) {\hat\psi}({\bf r}_{1}^{\prime})\rangle
=\int \d^{3}p_{1} \Bigl(n_{0}\delta({\bf p}_{1})+\frac{n(p_{1})}{(2\pi)^{3}}
\Bigr)
$$
$$
\times\exp\left[i{\bf p}_{1}({\bf r}_{1}^{\prime}-{\bf r}_{1})\right].
$$
By analogy with the above proof one can readily be convinced that the
distribution $w({\bf p},{\bf q})= w(\frac{{\bf p}_{1}-{\bf p}_{2}}{2},{\bf
p}_{1}+{\bf p}_{2})$ contains the following parts:
\begin{eqnarray}
1)\quad && n^{2}_{0}\delta({\bf p}_{1})\delta({\bf p}_{2}),
\label{wpqsing1}\\
2)\quad && n_{0}\delta({\bf p}_{1})\frac{n(p_{2})}{(2\pi)^{3}}
+n_{0}\delta({\bf p}_{2})\frac{n(p_{1})}{(2\pi)^{3}},
\label{wpqsing2}\\
3)\quad && \frac{n(p_{1})}{(2\pi)^{3}} \frac{n(p_{2})}{(2\pi)^{3}}.
\label{wpqsing3}
\end{eqnarray}
The expression (\ref{wpqsing1}) corresponds to condensate pair states, or
condensate-condensate pairs (both particle of a couple are in the
condensate); the terms (\ref{wpqsing2}), to condensate-supracondensate pairs;
and the last term (\ref{wpqsing3}), to supracondensate-supracondensate ones.
Thus, owing to the expressions (\ref{wpqsing1})-(\ref{wpqsing3}) we obtain
the natural classification of the ``dissociated" pair states.

We note that the total number of bound pairs $N_{b} \simeq V\sum_{i}\int
\d^{3}q\, w_{i}(q)$ is asymptotically proportional to $V^{1}$, \\while the
to\-tal num\-ber of ``dis\-so\-ci\-at\-ed" pa\-irs $N_{d} \simeq V^{2} \int
\d^{3} p\,\d^{3}q\; w({\bf p},{\bf q})=N^{2}$ is asymptotically proportional
to $V^{2}$.  From an intuitive point of view it is obvious, as a given
particle in a liquid can form bound states with $M$ particles
($M=\mbox{const}$ as $V\rightarrow \infty$), while it forms ``dissociated"
states with the other $N-1-M$ particles. So, we arrive at the following
asymptotic equation:
$$
N_{b}+N_{d}=N(N-1)\simeq N_{d}\simeq N^{2}.
$$

Diagonal elements of $F_{2}(\r{1},\r{2};\rp{1},\rp{2})$ depend on squ\-ared
absolute values of the PWF and determine the behaviour of the static pair
distribution function $g(r)$, which can be directly observed in scattering
experiments:
$$
g(r)=\frac{V^{2}}{N(N-1)}F_{2}(\r{},{\bf R};\r{},{\bf R})
$$
$$
=\frac{V^{2}}{N(N-1)}\sum_{\omega,{\bf q}}\frac{N_{\omega,{\bf q}}}{V}
|\psi_{\omega,{\bf q}}({\bf r})|^{2}.
$$
After the thermodynamic limit this expression takes a form
\begin{eqnarray}
g(r)= \frac{1}{n^{2}}\biggl(
&&\sum_{i}\int \d^{3}q\;w_{i}(q)
|\varphi_{{\bf q},i}(\r{})|^{2}
\nonumber \\
&&+\int \d^{3}p\,\d^{3}q\;w({\bf p},{\bf q})
|\varphi_{{\bf p},{\bf q}}(\r{})|^{2}
\biggr).
\label{gr}
\end{eqnarray}
One can easily be convinced that $g(r)\rightarrow 1$ as $r\rightarrow
\infty$. The behaviour of $g(r)$ for $r\rightarrow 0$ is determined by the
PWF at small distances. Because of repulsion, the probability of finding
three particles in a very small volume is much lower than that of finding two
particles in the same volume. Therefore, the short-range behaviour of two
particles is determined by the ordinary two-body Schr\"odinger equation with
the pairwise interaction $\Phi(r)$.  Thus, the PWF for short distances
between two particles is proportional to the usual wave function
$\varphi_{\bf p}^{(0)}(\r{})$ of the two-body problem~\cite{kimball}:
\begin{equation}
\varphi_{{\bf p},{\bf q}}(\r{})\to C\varphi_{\bf p}^{(0)}(\r{}),
\label{bcond1}
\end{equation}
when $r\to 0$. This implies that for a singular potential the PWF goes to
zero for $r\rightarrow 0$, and it follows that $g(r)\rightarrow 0$ as
$r\rightarrow 0$.

\section{Structure of pair correlation function and pair wave
functions for Bose liquid with condensate}
\label{3}

In order to examine the behaviour of the pair correlation function (\ref{f2})
with a condensate we will employ the procedure proposed first by
Bogoliubov~\cite{bog47} and lately justified with the help of the principle
of correlation weakening~\cite{bogquasi}: in the calculation of
averages~\cite{note1} of any number of the field operators we can substitute
the $C$-number $\sqrt{N_{0}}$ for the operators ${\hat a}_{0}$ and ${\hat
a}_{0}^{\dagger}$ involved in ${\hat\psi}(\r{})$ and
${\hat\psi}^{\dagger}(\r{})$. Here $N_{0}$ is the number of particles in the
condensate. It should be stressed that, after Bogoliubov, this substitution
gives {\it asymptotically exact values} of the averages in the limit
$V\rightarrow \infty$ and {\it does not imply any approximations}.  Indeed,
we have for any average that involves the operator ${\hat a}_{0}/\sqrt{V}$:
$$
\lim_{V\to\infty}\langle \frac{{\hat a}_{0}}{\sqrt{V}}\cdots{\hat\psi}({\bf r}')
\cdots\rangle
$$
$$
=\lim_{V\to\infty}\frac{1}{V}\int_{V}\d^{3}r\langle {\hat\psi}({\bf r})
\cdots{\hat\psi}({\bf r}')\cdots\rangle
$$
$$
=\lim_{V\to\infty}\frac{1}{V}\int_{V}\!\d^{3}r\langle{\hat\psi}({\bf r})\rangle
\langle\cdots{\hat\psi}({\bf r}')\cdots\rangle\!=\!
\sqrt{n_{0}}\langle\cdots{\hat\psi}({\bf r}')\cdots\rangle.
$$
Here we use the definition $\langle{\hat\psi}({\bf r})\rangle=
\langle{\hat\psi}^{\dagger}({\bf r})\rangle=\sqrt{n_{0}}$ ($n_{0}=N_{0}/V$ is
the density of the condensate) and the fact that the main contribution to the
value of the integral comes from infinitely large $r$. Thus, we can put
\begin{eqnarray}
{\hat\psi}(\r{})&=&\sqrt{n_{0}}+{\hat\vartheta}(\r{}),\nonumber \\
{\hat\psi}^{\dagger}(\r{})&=&\sqrt{n_{0}}+{\hat\vartheta}^{\dagger}(\r{}),
\label{psikrr}
\end{eqnarray}
where
\begin{eqnarray}
{\hat\vartheta}(\r{})&=&\frac{1}{\sqrt{V}}\sum_{{\bf p}\not=0}{\hat a}_{{\bf p}}
e^{i{\bf p}{\bf r}},\nonumber \\
{\hat\vartheta}^{\dagger}(\r{})&=&\frac{1}{\sqrt{V}}\sum_{{\bf p}\not=0}
{\hat a}^{\dagger}_{{\bf p}}e^{-i{\bf p}{\bf r}}
\label{thetakrr}
\end{eqnarray}
are supracondensate field operators. Substituting the expression
(\ref{psikrr}) into (\ref{f2}) and using the relation
$\langle{\hat\vartheta}({\bf r})\rangle=\langle{\hat\vartheta}^{\dagger }
({\bf r})\rangle=0$ we obtain
\end{multicols}\vspace*{-3mm}\noindent\rule[2mm]{87.36mm}{.1mm}%
\rule[2mm]{.1mm}{2mm}\vspace*{-2mm}
\begin{eqnarray}
F_{2}(\r{1},\r{2};\rp{1},\rp{2})=F^{(1)}_{2}(\r{1},\r{2};\rp{1},
\rp{2})+F^{(2)}_{2}(\r{1},\r{2};\rp{1},\rp{2})
+F^{(3)}_{2}(\r{1},\r{2};\rp{1},\rp{2})
+F^{(4)}_{2}(\r{1},\r{2};\rp{1},\rp{2}),
\label{f2tot}
\end{eqnarray}
where we introduce the notations
\begin{eqnarray}
F^{(1)}_{2}(\r{1},\r{2};\rp{1},\rp{2})&=&n_{0}^{2}+
n_{0}\langle{\hat\vartheta}^{\dagger}(\r{1}){\hat\vartheta}^{\dagger}(\r{2})
\rangle
+n_{0}\langle{\hat\vartheta}(\rp{2}){\hat\vartheta}(\rp{1})\rangle,
\label{f21}\\
F^{(2)}_{2}(\r{1},\r{2};\rp{1},\rp{2})&=&n_{0}\Bigl(
\langle{\hat\vartheta}^{\dagger}(\r{1}){\hat\vartheta}(\rp{1})\rangle+
\langle{\hat\vartheta}^{\dagger}(\r{1}){\hat\vartheta}(\rp{2})\rangle
+\langle{\hat\vartheta}^{\dagger}(\r{2}){\hat\vartheta}(\rp{2})\rangle
+\langle{\hat\vartheta}^{\dagger}(\r{2}){\hat\vartheta}(\rp{1})\rangle
\Bigr),
\label{f22}\\
F^{(3)}_{2}(\r{1},\r{2};\rp{1},\rp{2})&=&\sqrt{n_{0}}\Bigl(
\langle{\hat\vartheta}^{\dagger}(\r{2}){\hat\vartheta}(\rp{2}){\hat\vartheta}(\rp{1})\rangle
+\langle{\hat\vartheta}^{\dagger}(\r{1}){\hat\vartheta}(\rp{2}){\hat\vartheta}(\rp{1})\rangle
\nonumber \\
&&\phantom{\sqrt{n_{0}}\Bigl(}+\langle{\hat\vartheta}^{\dagger}(\r{1}){\hat\vartheta}^{\dagger}(\r{2}){\hat\vartheta}(\rp{2})
\rangle
+\langle{\hat\vartheta}^{\dagger}(\r{1}){\hat\vartheta}^{\dagger}(\r{2})
{\hat\vartheta}(\rp{1})
\rangle\Bigr),
\label{f23}\\
F^{(4)}_{2}(\r{1},\r{2};\rp{1},\rp{2})&=&
\langle{\hat\vartheta}^{\dagger}(\r{1}){\hat\vartheta}^{\dagger}(\r{2})
{\hat\vartheta}(\rp{2}){\hat\vartheta}(\rp{1})\rangle.\label{f24}
\end{eqnarray}
The expressions (\ref{f21})-(\ref{f24}) can be expanded in systems of
projectors (in the Hilbert space). For (\ref{f21})-(\ref{f23}) the
corresponding projectors can be expressed in an explicit form in terms of
averages of the creation and destruction Bose operators ${\hat a}_{{\bf p}}$
and ${\hat a}^{\dagger}_{{\bf p}}$. For (\ref{f21}) we have
$$
F^{(1)}_{2}(\r{1},\r{2};\rp{1},\rp{2})=n_{0}^{2}\varphi^{*}
(\r{1}-\r{2})
\varphi(\rp{1}-\rp{2})
-\langle{\hat\vartheta}^{\dagger}(\r{1}){\hat\vartheta}^{\dagger}(\r{2})\rangle
\langle{\hat\vartheta}(\rp{2}){\hat\vartheta}(\rp{1})\rangle,
$$
where
\begin{eqnarray}
\varphi(r)&=&1+\psi(r),\label{phir}\\
\psi(r)&=&\psi(\r{1}-\r{2})=\psi^{*}(\r{1}-\r{2})=
\langle{\hat\vartheta}(\r{1}){\hat\vartheta}(\r{2})\rangle/n_{0}
=\frac{1}{(2\pi)^{3}}\int \d^{3}k\;\psi(k)e^{i{\bf k}{\bf r}},
\label{psir}\\
\psi(k)&=&\langle {\hat a}_{{\bf k}}{\hat a}_{-{\bf k}}\rangle/n_{0}.
\label{psik}
\end{eqnarray}
In a similar manner we get
\begin{equation}
F_{2}^{(2)}(\r{},{\bf R};\rp{},{\bf R}^{\prime})=
\int \d^{3}p\,\d^{3}q\;w^{(1)}({\bf p},{\bf q})
\sqrt{2}\cos({\bf p}\r{})\sqrt{2}\cos({\bf p}\rp{})
\exp[i{{\bf q}}({\bf R}^{\prime }-{\bf R})].
\label{f22red}
\end{equation}
Here
\begin{equation}
w^{(1)}({\bf p},{\bf q})=2n_{0}\delta({\bf q}/2-{\bf p})
\frac{n({\bf q}/2+{\bf p})}{(2\pi)^{3}}
=2n_{0}\delta({\bf q}/2-{\bf p})\frac{n(q)}{(2\pi)^{3}}
\label{w1pq}
\end{equation}
is the momentum distribution for con\-den\-sate-sup\-ra\-con\-den\-sate
pairs corresponding to the expression (\ref{wpqsing2}).

Performing the Fourier transformations, we reduce $F_{2}^{(3)}$ to the
following form
\begin{eqnarray}
F_{2}^{(3)}(\r{},{\bf R};\rp{},{\bf R}^{\prime})=
\int \d^{3}p\,\d^{3}q\;w^{(1)}({\bf p},{\bf q})
\left(\psi_{{\bf p}}(\rp{})\sqrt{2}\cos({\bf p}\r{})+
\psi^{*}_{{\bf p}}(\r{})\sqrt{2}\cos({\bf p}\rp{})\right)
\times\exp[i{{\bf q}}({\bf R}^{\prime }-{\bf R})].
\label{f23red}
\end{eqnarray}
where
\begin{eqnarray}
\psi_{{\bf p}}(\r{})&&=
\frac{1}{(2\pi)^{3}}\int \d^{3}k\;\psi_{{\bf p}}({\bf k})
e^{i{\bf k}{\bf r}},
\label{psipr}\\
\psi_{{\bf p}}({\bf k})&&=
\psi_{{\bf p}}(-{\bf k})=\psi_{-{\bf p}}({\bf k})
=\sqrt{N_{0}}\frac{1}{\sqrt{2}n_{0}}
\frac{\langle {\hat a}^{\dagger}_{2{\bf p}}{\hat a}_{{\bf p}+{\bf k}}
{\hat a}_{{\bf p}-{\bf k}}\rangle}{n(2p)}.
\label{psipk}
\end{eqnarray}
The second equation in (\ref{psipk}) is a consequence of the invariance of
averages with respect to the transformation ${\hat a}_{{\bf k}}\to {\hat
a}_{-{\bf k}},\ {\hat a}^{\dagger}_{{\bf k}}\to {\hat a}^{\dagger}_{-{\bf
k}}$, as the Hamiltonian does not change when the transformation is
performed.  The factor $\sqrt{N_{0}}$ compensates the decrease of the average
value $\langle {\hat a}^{\dagger}_{2{\bf p}}{\hat a}_{{\bf p}+{\bf k}} {\hat
a}_{{\bf p}-{\bf k}}\rangle\sim V^{-1/2}$, so, the quantity $\psi_{{\bf
p}}({\bf k})$ is of order $1$ in the limit $V\rightarrow \infty$.

From (\ref{psipr}) and (\ref{psipk}) we obtain the symmetry properties
$$
\psi_{{\bf p}}({\bf r})=\psi_{{\bf p}}(-{\bf r})=
\psi_{-{\bf p}}({\bf r}).
$$
Summing the expressions (\ref{f22red}) and (\ref{f23red}) we get
$$
F_{2}^{(2)}(\r{},{\bf R};\rp{},{\bf R}^{\prime})+
F_{2}^{(3)}(\r{},{\bf R};\rp{},{\bf R}^{\prime})
=\int \d^{3}p\,\d^{3}q\;w^{(1)}({\bf p},{\bf q})
\varphi^{*}_{{\bf p}}(\r{})\varphi_{{\bf p}}(\rp{})
\exp[i{{\bf q}}({\bf R}^{\prime }-{\bf R})]
$$
$$
-\int \d^{3}q\;2n_{0}\frac{n(q)}{(2\pi)^{3}}
\psi^{*}_{{\bf q}/2}(\r{})\psi_{{\bf q}/2}(\rp{})
\exp[i{{\bf q}}({\bf R}^{\prime }-{\bf R})],
$$
where
\begin{equation}
\varphi_{{\bf p}}(\r{})=\varphi_{-{\bf p}}(\r{})=
\varphi_{{\bf p}}(-\r{})=\sqrt{2}\cos({\bf p}\r{})+
\psi_{{\bf p}}(\r{}),
\label{phipr}
\end{equation}
in accordance with (\ref{phisymm}).

Now it is not difficult to express (\ref{f2tot}) in a form similar to
(\ref{f2asymp})
\begin{eqnarray}
F_{2}(\r{},{\bf R};\rp{},{\bf R}^{\prime})=
n_{0}^{2}\varphi^{*}(r)\varphi(r')
+\int\!\d^{3}p\,\d^{3}q\;2n_{0}\delta({\bf q}/2-{\bf p})\frac{n(q)}{(2\pi)^{3}}
\varphi^{*}_{{\bf p}}(\r{})\varphi_{{\bf p}}(\rp{})
\exp[i{{\bf q}}({\bf R}^{\prime }-{\bf R})]
+{\widetilde F}_{2}(\r{},{\bf R};\rp{},{\bf R}^{\prime}),
\label{f2final}
\end{eqnarray}
where
\begin{eqnarray}
{\widetilde F}_{2}(\r{},{\bf R};\rp{},{\bf R}^{\prime})=
{\widetilde F}_{2}(\r{1},\r{2};\rp{1},\rp{2})
&=&F^{(4)}_{2}(\r{1},\r{2};\rp{1},\rp{2})-
\langle{\hat\vartheta}^{\dagger}(\r{1}){\hat\vartheta}^{\dagger}(\r{2})\rangle
\langle{\hat\vartheta}(\rp{2}){\hat\vartheta}(\rp{1})\rangle\nonumber \\
&&-\int \d^{3}q\;2n_{0}\frac{n(q)}{(2\pi)^{3}}
\psi^{*}_{{\bf q}/2}(\r{})\psi_{{\bf q}/2}(\rp{})
\exp[i{{\bf q}}({\bf R}^{\prime }-{\bf R})].
\label{f2tild}
\end{eqnarray}
\hfill\vspace*{-2mm}\rule[0mm]{.1mm}{2mm}%
\rule[2mm]{87.36mm}{.1mm}\vspace*{-3mm}\begin{multicols}{2}
As it is proved in Sec.~\ref{2}, the eigenvalues of $F_{2}(\r{},{\bf
R};\rp{},{\bf R}^{\prime})$ are given by Eqs.
(\ref{wpqsing1})-(\ref{wpqsing3}). So, one can conclude that the projectors
$\varphi^{*}(r)\varphi(r')$ and $\varphi^{*}_{{\bf q}/2}(\r{})\exp[-i{\bf
q}{\bf R}] \varphi_{{\bf q}/2}(\rp{}) \exp[i{\bf q}{\bf R}^{\prime}]$ are
nothing else but the eigenprojectors for the pair correlation function, and,
hence, $\varphi(r)$ and $\varphi_{{\bf q}/2}({\bf r})\exp[i{\bf q}{\bf R}]$
are the PWF for the Bose system below $T_{c}$.  The first term in
(\ref{f2final}) corresponds to pairs in the condensate; the second, to
condensate-supracondensate pairs; the last one, ${\widetilde F}_{2}$, to
pairs in the supracondensate, and, maybe, to bound pairs. The PWF
$\varphi(r)$ and $\varphi_{\bf p}({\bf r})$ given by expressions
(\ref{phir})-(\ref{psik}) and (\ref{psipr})-(\ref{phipr}) respectively are
normalized in accordance with (\ref{freenorm}).  The condensate PWF differs
essentially from $1$ at short distances and goes to $1$ when distances
between particles in a pair increase.  The situation is similar to the
scattering of two bare particles with zero momenta at infinity. The anomalous
averages $\langle{\hat\vartheta}^{\dagger}(\r{1})
{\hat\vartheta}^{\dagger}(\r{2})\rangle$ in PWF (\ref{phir}) have essentially
different physical interpretation in comparison with the fermion systems:
they determine the ``scattering" part of the condensate PWF, while for
fermion systems anomalous averages $\langle{\hat\psi}^{\dagger}(\r{1})
{\hat\psi}^{\dagger}(\r{2})\rangle$ themselves are PWF for the bound state of
pairs of particles with zero momentum~\cite{bogquasi,CSPRB}.

In the condensate-supracondensate PWF (\ref{phipr}) we drop the index
corresponding to the centre-of-mass momentum as ${\bf q}=\pm 2{\bf p}$.
This is a consequence of the $\delta$-function being present in the pair
momentum distribution (\ref{w1pq}). For the same reason in the
condensate-condensate PWF (\ref{phir}) the indices are omitted at all.

The functions $\psi(r)$ and $\psi_{{\bf p}}({\bf r})$ defined by Eqs.
(\ref{psir}) and (\ref{psipr}) are responsible for particle short-distance
correlations, therefore, owing to (\ref{bcond1}) for strongly singular
potentials we have
\begin{equation}
\psi(r=0)=-1,\quad\psi_{{\bf p}}({\bf r}=0)=-\sqrt{2},
\label{bcond}
\end{equation}
(see the discussion at the end of Sec. \ref{2}).

It should be emphasized that the pair correlation function for
supracondensate particles is ${\widetilde F}_{2}$ rather than $F_{2}^{(4)}$
defined by Eq. (\ref{f24}). Since the kernel (\ref{f2final}) is non-negative,
the kernel (\ref{f2tild}) is also non-negative.  Consequently, the last two
terms in the expression for $\widetilde {F}_{2}$ are reduced to zero by the
terms in $F_{2}^{(4)}$.  These terms look like those corresponding to bound
states of pairs. If the kernel $F_{2}^{(4)}$ written in form (\ref{f2asymp})
contains other terms associated with bound states, then the latter are also
contained in ${\widetilde F}_{2}$ and, consequently, in the final expression
(\ref{f2final}).  The distribution over supracondensate-supracondensate pair
states is given by the expression (\ref{wpqsing3}), and thus, we arrive at
the final formula for ${\widetilde F}_{2}$:
\begin{eqnarray}
{\widetilde F}_{2}&&(\r{},{\bf R};\rp{},{\bf R}^{\prime})
\nonumber \\
&&=\sum_{i}\int \d^{3}q\;w_{i}(q)
\varphi^{*}_{{\bf q},i}(\r{})\varphi_{{\bf q},i}(\rp{})
\exp[i{{\bf q}}({\bf R}^{\prime }-{\bf R})]\nonumber \\
&&\phantom{=}+\int \d^{3}p\,\d^{3}q\;
\frac{n({\bf q}/2+{\bf p})}{(2\pi)^{3}}
\frac{n({\bf q}/2-{\bf p})}{(2\pi)^{3}}
\nonumber \\
&&\phantom{=+\int}\times\varphi^{*}_{{\bf p},{\bf q}}(\r{})\varphi_{{\bf p},{\bf q}}
(\rp{})
\exp[i{{\bf q}}({\bf R}^{\prime }-{\bf R})],
\label{f2tildfin}
\end{eqnarray}
where the symmetry properties (\ref{phisymm}) are fulfilled.  Thus, in order
to answer the question about real bound states of pairs of particles, one
must carry out a detailed analysis taking into account all groups of the
terms (\ref{f21})-(\ref{f24}). We emphasize that we do not specify the size
of the bound pairs [the characteristic range of the PWF $\varphi_{{\bf
q},i}(\r{})$]. If it is much more than the mean distance between particles,
then, following Bogoliubov~\cite{bogquasi}, one may call these pairs
``quasi-molecules"~\cite{note8}. If the radius of the bound particle couples
is of the order of the mean distance between particles or, even, less then
one may speak about ordinary molecules. Note that ``quasi-molecules" are not
stable for a Bose system within the mean-field variational
treatment~\cite{nozieres} which is equivalent to the HFB approach. It is a
difficult question whether it is the case beyond the mean-field
approximation, so, {\it a priori} we cannot ignore existence of
``quasi-molecules".  The bound pair states can totally be a result of the
collective effects, as in the case of Fermi systems~\cite{CSPRB}. We note
that such situation can even be realized in the case of three
particles~\cite{baum}.

It is not difficult to see with the help of the Riemann theorem~\cite{Note6}
that the boundary condition (\ref{corr1}) is satisfied for (\ref{f2final}),
(\ref{f2tildfin}) in the limit (\ref{limit1}).

The expressions (\ref{f21})-(\ref{f24}) can be interpreted in the
conventional way, in terms of scattering processes for pairs of particles
$({\bf p}_{1},{\bf p}_{2})\rightarrow({\bf p}^{\prime}_{1},{\bf
p}^{\prime}_{2})$, where ${\bf p}_{1}$, ${\bf p}_{2}$, ${\bf p}^{\prime}_{1}$
and ${\bf p}^{\prime}_{2}$ are the particle momenta. For this purpose one
should calculate the averages in the momentum representation, i.e. make
Fourier transformation.  This procedure corresponds to expansion of the PWF
in plane waves.  Thus, the first term in (\ref{f21}) is related to ${\bf
p}_{1}={\bf p}_{2}={\bf p}^{\prime}_{1}={\bf p}^{\prime}_{2}=0$; the second
one, to ${\bf p}_{1},\ {\bf p}_{2}\not=0$ and ${\bf p}^{\prime}_{1}={\bf
p}^{\prime}_{2}=0$; and so on. The terms in (\ref{f23}) correspond to
processes when one of the four momenta is zero. These processes become
dominant in the vicinity of $T_{c}$~\cite{note7}, when $N_{0}/N\ll 1$. In
this case $\left(\frac{N_{0}}{N}\right)^{2}\ll\frac{N_{0}}{N}\ll
\sqrt{\frac{N_{0}}{N}}$, so one can expect that $F^{(3)}_{2}$ is important in
the expression (\ref{f2tot}). Taking into account that the thermodynamic
potential is tightly connected with the correlation function, one can assume
that the terms in (\ref{f23}) are important to obtain correct thermodynamics
at those temperatures. Recall that these are precisely the terms being
responsible for the short-range correlations for the
condensate-supracondensate pairs.

\section{Pair wave functions in Hartree-Fock-Bogoliubov and Bogoliubov
models}
\label{3a}

Let us consider the HFB approach from the point of view of PWF. In the HFB
method (see the detailed discussion in Ref.~\cite{hohmartin}), a Hamiltonian
is approximated by a quadratic form of the Bose operators ${\hat\alpha}_{{\bf
p}}^{\dagger}$ and ${\hat\alpha}_{{\bf p}}$ connected with initial operators
${\hat a}_{{\bf p}}^{\dagger}$ and ${\hat a}_{{\bf p}}$ by the canonical
Bogoliubov transformations.  Consequently, all averages can be calculated
with the Wick-Bloch-De Dominicis theorem~\cite{bloch}. Thus, we have
\end{multicols}\vspace*{-3mm}\noindent\rule[2mm]{87.36mm}{.1mm}%
\rule[2mm]{.1mm}{2mm}\vspace*{-2mm}
\begin{eqnarray}
F^{(3)}_{2}(\r{1},\r{2};\rp{1},\rp{2})&=&0,
\label{f32hfb}\\
F^{(4)}_{2}(\r{1},\r{2};\rp{1},\rp{2})&=&
\langle{\hat\vartheta}^{\dagger}(\r{1}){\hat\vartheta}(\rp{1})\rangle
\langle{\hat\vartheta}^{\dagger}(\r{2}){\hat\vartheta}(\rp{2})\rangle
+\langle{\hat\vartheta}^{\dagger}(\r{1}){\hat\vartheta}(\rp{2})\rangle
\langle{\hat\vartheta}^{\dagger}(\r{2}){\hat\vartheta}(\rp{1})\rangle
+\langle{\hat\vartheta}^{\dagger}(\r{1}){\hat\vartheta}^{\dagger}(\r{2})\rangle
\langle{\hat\vartheta}(\rp{2}){\hat\vartheta}(\rp{1})\rangle.
\label{f24hfb}
\end{eqnarray}
The correlation function (\ref{f2tot}) can easily be rewritten in
the following form
\begin{eqnarray}
F_{2}(\r{},{\bf R};\rp{},{\bf R}^{\prime})=
n_{0}^{2}\varphi^{*}(r)\varphi(r')
+\int &&\d^{3}p\,\d^{3}q\;
\Bigl[2n_{0}\delta({\bf q}/2-{\bf p})\frac{n(q)}{(2\pi)^{3}}
+\frac{n({\bf q}/2+{\bf p})}{(2\pi)^{3}}
\frac{n({\bf q}/2-{\bf p})}{(2\pi)^{3}}\Bigr]\nonumber \\
&&\times\sqrt{2}\cos({\bf p}\r{})\sqrt{2}\cos({\bf p}\rp{})
\exp[i{{\bf q}}({\bf R}^{\prime }-{\bf R})],
\label{f2hfb}
\end{eqnarray}
where $\varphi(r)$ is related to the averages via (\ref{phir}). In so doing,
$n(p)$ and $\psi(k)$ are expressed in terms of Hamiltonian parameters in a
self-consistent manner. From (\ref{f2hfb}) one can see that there are no
bound states of particles in the framework of the HFB approach. We note that
the expression (\ref{f2hfb}) corresponds to the ideal Bose gas when
$\varphi(r)=1$ and $n(p)$ is the Bose-Einstein distribution with a zero
chemical potential. One can see that in this case all PWF are symmetrized
plane waves, as it is to be expected.

In the HFB method, only the condensate PWF differs from the plane wave. Since
PWF are intimately related to the excitation spectrum, one can assume that it
is a rough approximation, which has led to the unphysical gap in the
single-particle excitation energy.  As Eq. (\ref{f32hfb}) is fulfilled for
all temperatures, then the HFB model cannot correctly describe the
thermodynamic behaviour of the system in the vicinity of $T_{c}$ ($T<T_{c}$).
It should be pointed out that the expression (\ref{f2hfb}) is valid provided
$n_{0}\not=0$. If $n_{0}=0$, but $\langle {\hat\vartheta}^{\dagger}(\r{1})
{\hat\vartheta}^{\dagger}(\r{2}) \rangle \not=0$, as in the theory of Valatin
and Butler~\cite{val} (see the careful analysis by Nozi\`eres and Saint
James~\cite{nozieres}), then $F_{2}^{(1)}=F_{2}^{(2)}=F_{2}^{(3)}=0$ and
$F_{2}^{(4)}$ is determined by (\ref{f24hfb}). In this case the anomalous
averages really correspond to the bound pairs of particles with $q=0$:
$$
F_{2}(\r{},{\bf R};\rp{},{\bf R}^{\prime})=
\rho_{0}\varphi^{*}(r)\varphi(r')
+\int \d^{3}p\,\d^{3}q\;
\frac{n({\bf q}/2+{\bf p})}{(2\pi)^{3}}
\frac{n({\bf q}/2-{\bf p})}{(2\pi)^{3}}
\sqrt{2}\cos({\bf p}\r{})\sqrt{2}\cos({\bf p}\rp{})
\exp[i{{\bf q}}({\bf R}^{\prime }-{\bf R})],
$$
\noindent\hspace{22pc}\vspace*{-2mm}\rule[0mm]{.1mm}{2mm}%
\rule[2mm]{86.36mm}{.1mm}\vspace*{-3mm}\begin{multicols}{2}
where $\rho_{0}$ is the density of the bound pairs
$$
\rho_{0}=\int \d^{3}r_{1}\;
|\langle{\hat\vartheta}(\r{1}){\hat\vartheta}(\r{2})\rangle|^{2},
$$
and
$\varphi(\r{1}-\r{2})= \langle {\hat\vartheta}(\r{1}) {\hat\vartheta}(\r{2})
\rangle /\sqrt{\rho_{0}}$ is PWF normalized in accordance with
(\ref{svnorm})~\cite{note10}.

The Bogoliubov theory of weakly imperfect Bose gas is a particular case of
the HFB theory when the condensate depletion is small:  $(n-n_{0})/n\ll 1$.
It can be obtained directly from (\ref{f2tot}) if we neglect terms of the
third and forth orders in ${\hat\vartheta}$ and
${\hat\vartheta}^{\dagger}$~\cite{note2}
\begin{eqnarray}
F_{2}&&(\r{},{\bf R};\rp{},{\bf R}^{\prime})
\nonumber \\
=&&n_{0}^{2}\varphi^{*}(r)\varphi(r')
+\int \d^{3}p\,\d^{3}q\;
2n_{0}\delta({\bf q}/2-{\bf p})\frac{n(q)}{(2\pi)^{3}}\nonumber \\
&&\times\sqrt{2}\cos({\bf p}\r{})\sqrt{2}\cos({\bf p}\rp{})
\exp[i{{\bf q}}({\bf R}^{\prime }-{\bf R})].
\label{f2bog}
\end{eqnarray}
Comparing the exact expressions (\ref{f2final}) and (\ref{f2tildfin}) with
Eq.~(\ref{f2bog}) one can reveal that the Bogoliubov model involves the
following approximations.  First, the supracondensate-supracondensate part is
omitted, which is completely justified as the condensate depletion is small.
Second, the condensate-supracondensate PWF are symmetrized plane waves
[$\psi_{{\bf p}}(\r{})=0$ in Eq. (\ref{phipr})], hence, the boundary
conditions (\ref{bcond}) for a hard-core potential cannot be satisfied by any
choice of possible parameters. In particular, the change of the ``bare"
potential by the effective one leaves this property of the model unaltered.
The approximation $\varphi_{\bf p}(\r{})=\sqrt{2} \cos({\bf p}{\bf r})$
implies that the Bogoliubov model is nothing else but the ideal gas
approximation for the condensate-supracondensate PWF, while the scattering
part of the condensate-condensate PWF (\ref{psir}) is not zero but small and
can be evaluated within the Born approximation for the in-medium
PWF~\cite{CSPRE,CSJPS,CSPL}.  Therefore, the Bogoliubov model does not
properly take into account the short-range correlations of bosons. So, in the
framework of this model we obviously obtain the divergency when evaluating
the mean energy for the ``bare" hard-core potential due to the infinite
contribution of the condensate-supracondensate ``channel" involved in the
representation (\ref{gr}) to the mean interaction energy per particle
(\ref{uaver}). The hard-sphere model can be treated as the Bogoliubov model
with the replacement $\Phi(r)\to\delta({\bf r})4\pi a/m$.  Even after this
substitution we face another divergency. The point is that the Bogoliubov
model involves, in the implicit form, the next-to-Born term for the
scattering amplitude: $a_{1}=-b$, $b=m/[4\pi(2\pi)^3]
\int\d^3k\,\Phi^{2}(k)/(2T_k)$, where $T_{k}=k^{2}/(2m)$. The hard-sphere
replacement implies that $\Phi(k)=4\pi a/m=\mbox{\rm const}$, which leads to
the divergency associated with the parameter $b$, and, hence, to divergent
the mean energy.  A cure for this difficulty is to replace the Bogoliubov
expression (\ref{f2bog}) by the exact formula (\ref{f2final}) in which the
term $\widetilde{F}_{2}$ can be neglected due to a small condensate
depletion:
\begin{eqnarray}
&&F_{2}(\r{},{\bf R};\rp{},{\bf R}^{\prime})=
n_0^2\varphi^*(r)\varphi(r')
\nonumber \\
&&+2n_0\int\!\frac{\d^3q}{(2\pi)^3}n(q)\varphi_{{\bf q}/2}^*({\bf r})
 \varphi_{{\bf q}/2}({\bf r}')
\exp[i{\bf q}({\bf R}'-{\bf R})].
\label{f2ans}
\end{eqnarray}
In so doing the PWF $\varphi(r)$ and $\varphi_{\bf p}({\bf r})$ should be
determined in a self-consistent manner which provides the boundary conditions
(\ref{bcond1}) [and, hence, (\ref{bcond})]. Along this line with the help of
the variational procedure we can derive the following non-linear equation:
\begin{equation}
U(k)=\Phi(k)
-\frac{1}{2}\int\frac{\d^3q}{(2\pi)^3}\frac{\Phi(|{\bf k}
-{\bf q}|)U(q)}{\sqrt{\widetilde{T}^2_q+
                                  2n\widetilde{T}_q U(q)}}
\label{Uk}
\end{equation}
valid at zero temperature and small densities~\cite{CSPRE,CSJPS}.  Here for
the function $\widetilde{T}_{q}$ we can put, within a good accuracy,
$\widetilde{T}_{q}\simeq T_{q}=q^{2}/(2m)$, $\Phi(k)$ is the Fourier
transform of the potential $\Phi(r)$, and
$U(k)=\int\d^{3}r\,\varphi(r)\Phi(r) \exp(-i{\bf k}{\bf r})$ can be treated
as a scattering amplitude in a medium. In the limit $n\to0$ Eq. (\ref{Uk}) is
reduced to the ordinary Lippmann-Schwinger equation for the scattering
amplitude which corresponds to the zero scattering momentum. Note that Eq.
(\ref{Uk}) looks like an equation for the many-body $t$-matrix $\Gamma({\bf
k},{\bf k}',{\bf q};\omega)$ at $\omega=0$ (see, e.g. the recent
review~\cite{shigrif}, Sec. 4).  However, there exist essential differences.
First, in general, the equation for the $t$-matrix is frequency dependent in
contrast to Eq.~(\ref{Uk}). Second, $\Gamma({\bf k},{\bf k}',{\bf
q};\omega=0)$ can not be associated with in-medium scattering amplitude
$U_{{\bf k}',{\bf q}}({\bf k})= \int\d^{3}r\,\varphi_{{\bf k}',{\bf q}}({\bf
r}) \Phi(r) \exp(-i{\bf k}{\bf r})$. In particular, the
condensate-supracondensate PWF (\ref{psipr})-(\ref{phipr}) is characterized
by the index ${\bf k}'=\pm{\bf q}/2$ and expressed via ``triple" averages
(\ref{psipk}) which are completely neglected in the many-body $t$-matrix
approximation as well as in the Bogoliubov, HFB and Popov
ones~\cite{shigrif}. Thus, one cannot associate the PWF $\varphi_{{\bf
k}',{\bf q}}({\bf r})$ with the function $\chi_{{\bf k}',{\bf
q};\omega=0}({\bf r})$ whose Fourier transform is usually defined as
$\Gamma({\bf k},{\bf k}',{\bf q};\omega)=1/(2\pi)^{3}
\int\d^{3}p\,\Phi(p)\chi_{{\bf k}',{\bf q};\omega}({\bf k}-{\bf p})$ [see
Ref.~\cite{shigrif}, Eq. (A.23)]. The latter is often called ``effective wave
function in a medium".  Third, Eq.~(\ref{Uk}) is obtained by means of {\it
variational} method {\it beyond} a mean-field
approximation~\cite{CSPRE,CSJPS}, while an equation for $t$-matrix is usually
derived by summing a certain set of the diagrams. For this reason the
relation $\widetilde{T}_{q}\simeq T_{q}=q^{2}/(2m)$ is an approximation for
Eq.~(\ref{Uk}) while the $t$-matrix formalism deals exactly with $T_{q}$.
Thus, the relation between the $t$-matrix and PWF formalism is rather subtle
and beyond the scope of this paper. We only note that our definition of the
PWF as the eigenfunctions of the 2-matrix is consistent with that of quantum
mechanics. Namely, we operate with the functions which determine the average
energy of the interaction per particle (\ref{uaver}) by means of the
representations (\ref{gr}) in accordance with the formalism of quantum
statistical mechanics. While the physical sence of the functions $\chi_{{\bf
k}',{\bf q};\omega=0}({\bf r})$ is rather unclear.

At sufficiently large momentum $k$ the main contribution in the integral in
the right-hand side of Eq. (\ref{Uk}) comes from the large momenta $q$. As
$\lim_{q\to\infty}U(q)/\widetilde{T}_{q}=0$, for $k\to\infty$ Eq. (\ref{Uk})
is also reduced to the two-body Lippmann-Schwinger equation, which leads to
the boundary condition (\ref{bcond1}). However, it follows from Eq.
(\ref{Uk}) that at finite density $n$ $\psi(k)\propto 1/k$ at small $k$, and,
hence, $\psi(r) \propto 1/r^{2}$ at $r\to\infty$, in contrast to the two-body
problem which implies $\psi(r)\propto 1/r$. Thus, peculiar overscreening
takes place for the condensate-condensate PWF.  Details concerning Eq.
(\ref{Uk}) can be found in Refs.~\cite{CSPRE,CSJPS,CSPRE2}.

With the help of the formalism of the PWF one can obtain the exact
relationship between the chemical potential and the PWF (\ref{phir}) and
(\ref{phipr}) in the presence of the Bose-Einstein condensate.  This
relationship is of special interest, for it allows us to see in an
explicit form how the renormalization of an interatomic interaction takes
place as applied to the Gross-Pitaevskii equation, at least in the
homogeneous case. Let us consider at first an inhomogeneous Bose system with
the Hamiltonian
\end{multicols}\vspace*{-3mm}\noindent\rule[2mm]{87.36mm}{.1mm}%
\rule[2mm]{.1mm}{2mm}\vspace*{-2mm}
\begin{equation}
\widehat H=\int\d^{3}r\,{\hat\psi}^{\dagger}({\bf r},t)
\Bigl(-\frac{\nabla^{2}}{2m}+V_{ext}({\bf r})\Bigr){\hat\psi}({\bf r},t)+
\int\d^{3}r\d^{3}r'\,\Phi(|{\bf r}-{\bf r}'|)
{\hat\psi}^{\dagger}({\bf r},t){\hat\psi}^{\dagger}({\bf r}',t)
{\hat\psi}({\bf r}',t){\hat\psi}({\bf r},t),
\label{ham}
\end{equation}
where the potential $V_{ext}({\bf r})$ corresponds to the external forces,
and $\Phi(r)$ is the ``bare" interaction potential. Note that the Bose field
operators used in Secs.~\ref{2} and \ref{3} are related to the zero time:
$\psi({\bf r})=\psi({\bf r},t=0)$, $\psi^{\dagger}({\bf r}) =
\psi^{\dagger}({\bf r},t=0)$.  Using the Heisenberg equation of motion with
the Hamiltonian (\ref{ham}) for the Bose field operator ${\hat\psi}({\bf r},t)$
and taking average values we get
\begin{equation}
i\frac{\partial \phi({\bf r},t)}{\partial t}=
\Bigl(-\frac{\nabla^{2}}{2m}+V_{ext}({\bf r})\Bigr)\phi({\bf r},t)+
\int\d^{3}r'\,\Phi(|{\bf r}-{\bf r}'|)
\langle{\hat\psi}^{\dagger}({\bf r}',t){\hat\psi}({\bf r}',t)
{\hat\psi}({\bf r},t)\rangle,
\label{psi0}
\end{equation}
where $\phi({\bf r},t)=\langle{\hat\psi}({\bf r},t)\rangle$ [and,
respectively, $\phi^{*}({\bf r},t)=\langle{\hat\psi}^{\dagger}({\bf
r},t)\rangle$].  On the other hand, one can make use of the Bogoliubov's
procedure of ``extracting" the $c$-number parts of the Bose field operators
in the Hamiltonian (\ref{ham}) (see discussion in Sec.~\ref{3}):
\begin{eqnarray}
{\hat\psi}(\r{},t)&=&\phi({\bf r},t)+{\hat\vartheta}(\r{},t),
\nonumber \\
{\hat\psi}^{\dagger}(\r{},t)&=&\phi^{*}({\bf r},t)+
{\hat\vartheta}^{\dagger}(\r{},t).
\label{psikrr1}
\end{eqnarray}
Thus, in the stationary equilibrium case the grand canonical potential
$\Omega$ depends on the $c$-number complex parameter $\phi({\bf r})$ that can
be considered as the order parameter. So, the equilibrium value of the grand
canonical potential $\Omega=\Omega(T,V,\mu;\{\phi({\bf r})\},\{\phi^{*}({\bf
r})\})$ corresponds to a minimum with respect to the order parameter (the
parameter $\mu$ stands for the chemical potential).  Using the well-known
expression for an infinitesimal change of the potential
$\delta\Omega=\langle\delta(\widehat H-\mu \widehat N)\rangle$ one can obtain
from the condition $\delta\Omega/\delta\phi({\bf r})=
\delta\Omega/\delta\phi^{*}({\bf r})=0$:
\begin{equation}
\mu\phi({\bf r})=
\Bigl(-\frac{\nabla^{2}}{2m}+V_{ext}({\bf r})\Bigr)\phi({\bf r})+
\int\d^{3}r'\,\Phi(|{\bf r}-{\bf r}'|)
\langle{\hat\psi}^{\dagger}({\bf r}'){\hat\psi}({\bf r}')
{\hat\psi}({\bf r})\rangle.
\label{statpsi0}
\end{equation}
It is not difficult to see that Eq.~(\ref{statpsi0}) is nothing but the
stationary form of Eq.~(\ref{psi0}) corresponding to the solution $\phi({\bf
r},t)=\phi({\bf r})\exp(-i\mu t)$ because in this case we have for the
time-dependent ``triple" averages
\begin{equation}
\langle{\hat\psi}^{\dagger}({\bf r}_{2},t){\hat\psi}({\bf r}'_{2},t)
{\hat\psi}({\bf r}'_{1},t)\rangle=\exp(-i\mu t)
\langle{\hat\psi}^{\dagger}({\bf r}_{2}){\hat\psi}({\bf r}'_{2})
{\hat\psi}({\bf r}'_{1})\rangle.
\label{timeevol}
\end{equation}
Indeed, the principle of the correlation weakening requires that at
$|{\bf r}_{1}|\to\infty$
$$
\langle{\hat\psi}^{\dagger}({\bf r}_{1}){\hat\psi}^{\dagger}({\bf r}_{2})
{\hat\psi}({\bf r}'_{2}){\hat\psi}({\bf r}'_{1})\rangle=
\langle{\hat\psi}^{\dagger}({\bf r}_{1},t){\hat\psi}^{\dagger}({\bf r}_{2},t)
{\hat\psi}({\bf r}'_{2},t){\hat\psi}({\bf r}'_{1},t)\rangle\to
$$
$$
\langle{\hat\psi}^{\dagger}({\bf r}_{1},t)\rangle
\langle{\hat\psi}^{\dagger}({\bf r}_{2},t)
{\hat\psi}({\bf r}'_{2},t){\hat\psi}({\bf r}'_{1},t)\rangle=
\langle{\hat\psi}^{\dagger}({\bf r}_{1})\rangle
\langle{\hat\psi}^{\dagger}({\bf r}_{2})
{\hat\psi}({\bf r}'_{2}){\hat\psi}({\bf r}'_{1})\rangle.
$$
\begin{multicols}{2}
The relation (\ref{timeevol}) follows from the last equation as
$\langle{\hat\psi}^{\dagger}({\bf r}_{1})\rangle= \phi^{*}({\bf r}_{1})$ and
$\langle{\hat\psi}^{\dagger}({\bf r}_{1},t)\rangle= \phi^{*}({\bf
r}_{1})\exp(i\mu t)$.  The stationary Gross-Pitaevskii equation can be
obtained from the exact equation (\ref{statpsi0}) with the help of the
replacement:
\begin{eqnarray}
\Phi(r)&\to&\frac{4\pi a}{m}\delta({\bf r}),
\label{replPhi} \\
\langle{\hat\psi}^{\dagger}({\bf r}'){\hat\psi}({\bf r}')
{\hat\psi}({\bf r})\rangle&\to&
\phi^{*}({\bf r}')\phi({\bf r}')
\phi({\bf r}),
\label{replpsi}
\end{eqnarray}
where $a$ is the scattering length which can be defined with the help of the
``bare" scattering  amplitude: $4\pi a/m=U^{(0)}(0)$ [$U^{(0)}(k)$ is the
solution of Eq.~(\ref{Uk}) at $n=0$ and $\widetilde{T}_{q} =
T_{q}=q^{2}/(2m)$]. However, within that approach it is not clear how the
renormalization (\ref{replPhi}) takes place.  The most transperent way of
understanding this replacement in the homogeneous case is based on the PWF
formalism.  In this situation $V_{ext}({\bf r})=0$,
$\phi=\sqrt{n_{0}}\exp(i\chi)$, and we can put the phase $\chi$ equal to
zero. So, for the homogeneous system the exact equation (\ref{statpsi0})
reads
\begin{equation}
\mu=\frac{1}{\sqrt{n_0}}\int\d^3r'\,\Phi(|{\bf r}-{\bf r}'|)
\langle{\hat\psi}^{\dagger}({\bf r}'){\hat\psi}({\bf r}')
{\hat\psi}({\bf r})\rangle.
\label{mu}
\end{equation}
This equation has been obtained in the momentum representation in the
Bogoliubov's paper~\cite{bogquasi} (see also Ref.~\cite{string}). Using the
specific expressions for the scattering parts of the condensate-condensate
and supracondensate-condensate PWF given by Eqs. (\ref{psik}) and
(\ref{psipk}), one can represent Eq.~(\ref{mu}) in the following form:
\begin{equation}
\mu=n_0 U(0)+
\sqrt{2}\int\frac{\d^3q}{(2\pi)^3}n(q)
U_{{\bf q}/2}({\bf q}/2),
\label{mu1}
\end{equation}
here $U_{{\bf p}}({\bf k})= \int\d^3r\,\varphi_{\bf p}({\bf r}) \Phi(r)
\exp(-i{\bf k}{\bf r})$.  Equation (\ref{mu1}) is the exact representation
for the chemical potential via the condensate-condensate and
condensate-supracondensate PWF and the occupation numbers $n(k)$. It can be
used in order to obtain the density expansion for $\mu$ at $n\to 0$. With the
help of the expression for $n(k)$ established within the
variational scheme mentioned above~\cite{CSPRE,CSJPS},
\begin{equation}
n(k)=\frac{1}{2}\Biggl(\frac{\widetilde{T}_k+n U(k)}
{\sqrt{\widetilde{T}_k^2+2n\widetilde{T}_k U(k)}}
-1\Biggr),
\label{nk}
\end{equation}
one can derive for the condensate density
\begin{eqnarray}
n_0&=&n\left(1-\int \frac{\d^{3}k}{(2\pi)^{3}} \frac{n(k)}{n}\right)
\nonumber \\
   &=&n\left(1-\frac{8}{3\sqrt{\pi}}\sqrt{na^3} + \cdots\right),
\label{depletion}
\end{eqnarray}
where the substitution ${\bf k}={\bf k}'\sqrt{2mn}$ in the integral is
employed. Besides, at $n\to 0$ from Eq. (\ref{Uk}) one can obtain the
density correction for the in-medium scattering
amplitude~\cite{CSEPJB,CSJPS,CSPRE2}
\begin{equation}
U(0) = U^{(0)}(0)\Bigl(1 + \frac{8}{\sqrt{\pi}}\sqrt{na^3}+ \cdots \Bigr).
\label{correction}
\end{equation}
Using Eqs. (\ref{depletion}) and (\ref{correction}) and making the ``scaling"
substitution ${\bf q}={\bf q}'\sqrt{2mn}$ in the integral in Eq. (\ref{mu1}),
we arrive at the familiar density expansion~\cite{brueck,lee,bel}
\begin{equation}
\mu=nU^{(0)}(0)\left(1+\frac{32}{3\sqrt{\pi}}
\sqrt{n a^3}+\cdots\right).
\label{mudens}
\end{equation}
In so doing, the explicit formula for the condensate-supracondensate PWF
$\varphi_{\bf p}({\bf r})$ are needless, but only the relation $\lim_{{\bf
p}\to 0}\varphi_{\bf p}({\bf r})=\sqrt{2}\varphi(r)$ is of use. The factor
$\sqrt{2}$ appears as due to the normalization condition (\ref{freenorm}).

Now it is not difficult to see that in the homogeneous case the
Gross-Pitaevskii equation corresponds to the simplest Hartree approximation
with the renormalized potential (\ref{replPhi}):
\begin{equation}
\mu\simeq n_0 U^{(0)}(0).
\label{mupit}
\end{equation}
The range of validity of this approximation is determined by the
next-to-leading terms in Eqs.~(\ref{depletion})-(\ref{mudens}) and readily
expressed as $\sqrt{na^{3}}\ll 1$. In the paper of Prokakis et
al.~\cite{burnett} the criterion for validity of the Gross-Pitaevskii
equation is formulated as
\begin{equation}
n\frac{4\pi\hbar^{2}a}{m}\ll \hbar\omega,
\label{critburn}
\end{equation}
where the frequency $\omega$ characterizes the harmonic trap potential. In
the homogeneous case ($\omega=0$) it cannot be fulfilled, while the
Gross-Pitaevskii approximation (\ref{replPhi}), (\ref{replpsi}) is valid, if
the condition $\sqrt{na^{3}}\ll 1$ is realized. Thus, the criterion
(\ref{critburn}) appears to be refined.

\section{Summary}
\label{4}

The reduced density matrix of the second order (\ref{rho2vtor}) is a
fundamental characteristic of a many-particle system, its eigenfunctions,
PWF, being the pure states of two particles, selected in an arbitrary way,
and its eigenvalues being the probabilities of finding a couple in those
states. Thus, the properties of the $2$-matrix are of interest in itself and
from the point of view of various applications as well.

With the Bogoliubov principle of correlation weakening we express the
momentum distribution of ``dissociated" pair states in terms of the
single-particle occupation numbers [see the expressions
(\ref{wpqsing1})-(\ref{wpqsing3})]. This allows us to represent the pair
correlation function (\ref{f2}) in the form (\ref{f2final}),
(\ref{f2tildfin}), where the condensate-condensate and
condensate-supracondensate PWF are explicitly given by Eqs.
(\ref{phir})-(\ref{psik}) and (\ref{psipr})-(\ref{phipr}), respectively. It
should be stressed that this representation is exact and do not implies any
mean-field approximation as well as various model assumptions since it is
solely based on the Bogoliubov's principle of the correlation weakening.  The
transparent physical interpretation of the anomalous averages (\ref{psir})
and (\ref{psipr}) as the ``scattering parts" of the corresponding PWF is
proposed. The boundary conditions (\ref{bcond}) are obtained in the case of
hard-core, or strongly singular, potentials. The condensate-supracondensate
PWF is determined by the ``triple" average (\ref{psipk}). The importance of
these averages is emphasized in Ref.~\cite{burnett} for the effects of
condensate--exited-state interactions in the evolution of condensate mean
field, which are in accordance with our results. In the
textbook~\cite{griffin} the crucial role of the contribution of the ``triple"
averages is discussed in the context of ``dielectric formalism" approach.
Those averages are not zero in so called Beliaev-Popov approximation
discussed recently in~\cite{shigrif}.

Thus, the anomalous averages
$\langle{\hat\vartheta}(\r{1}){\hat\vartheta}(\r{2})\rangle$ are responsible
for the spatial correlations of two particles in the condensate, and do not
imply that bound states of pairs of particles exist; the terms (\ref{f23})
are found to be responsible for the spatial correlations of
condensate-supracondensate pairs of particles and can be expected to be
important in thermodynamics near the critical temperature $T_{c}$. In the HFB
approach, the PWF are symmetrized plane waves except for the condensate pairs
provided the condensate density is not equal to zero.

The formalism of PWF is useful to construct various approximations which take
into account both short-range and long-range spatial correlations of
particles in a self-consistent manner. This is due to the fact that all the
pair wave functions in the expansion (\ref{f2final}), (\ref{f2tildfin}) are
``regular channels" from the point of view of its behaviour at small $r$.
Along this line one can obtain the system of non-linear
integro-differential equations for PWF, which in the weak-coupling
approximation leads to the Bogoliubov model of the weakly interacting Bose
gas~\cite{CSPL}.  For a dilute Bose gas the non-linear integral equation
(\ref{Uk}) that takes into account both short- and long-range spatial
correlations can be obtained with the help of variational method beyond a
mean-field approach~\cite{CSEPJB,CSJPS,CSPRE2}. The exact relationship between PWF
and the chemical potential (\ref{mu1}) allows us to obtain the density
expansion for the chemical potential (\ref{mudens}) in the strong-coupling
regime (Bose gas is dilute, but the potential is strongly singular as the
Lennard-Jones one). In this regime one can derive the expansion for the mean
energy per particle in powers of the boson density $n$, the famous
results~\cite{lee,bel} being reproduced without any divergencies which are
inherent in the pseudopotential approximation~\cite{CSEPJB,CSJPS,CSPRE2}. Thus,
this expansion can be readily obtained from Eq.~(\ref{mudens}) with the
thermodynamic relation $\mu=\partial (\varepsilon(n) n)/\partial n$:
$$
\varepsilon=\frac{1}{n}\int\limits_{0}^{n}\d n'\,\mu(n')
=\frac{2\pi\hbar^2 a n}{m}
\Bigl(1+\frac{128}{15\sqrt{\pi}}\sqrt{na^3}+\cdots\Bigr).
$$
Moreover, in the framework of our approach we can obtain by {\it direct}
calculation the interaction energy per particle~(\ref{uaver}) (as well as the
kinetic one)~\cite{CSEPJB}. However, this is impossible within the
effective-potential approach of Refs.~\cite{brueck,lee,bel}, which obviously
indicates the incorrect behaviour of the pair distribution function $g(r)$
(see the detailed discussions in Refs.~\cite{CSEPJB,CSPRE2}).

It is worth noting that the concept of PWF is helpful not only in the case of
the Bose systems but also in the situation of the Fermi ones.  Indeed,
employing this formalism one can easily prove that below the temperature of
the superconducting phase transition there always exist the bound states of
fermion pairs beyond the pair condensate~\cite{CSPRB}.

The author would like to thank A.~A.~Shanenko for fruitful discussions.


\end{multicols}

\begin{references}
\bibitem{note3} We mean the behaviour of these functions after the
limit $V\rightarrow \infty$.
\bibitem{note1} Bogoliubov introduced the notion ``quasi-averages" for
systems with spontaneously broken symmetry. So, by the notion ``averages" we
mean ``quasi-averages".
\bibitem{note10} As it is shown by P.~Nozi\`eres and D.~Saint
James~\cite{nozieres}, bound pair states can only exist for a system of
bosons with an internal spin structure when the two-body interaction $\Phi$
depends on the spins of the particles. We do not consider the spin of the
particles for the sake of simplicity, though we can easily generalize the
concept of the PWF for this case~\cite{bogquasi,note4,CSPRB}.
\bibitem{note2}In the Bogoliubov theory, $F^{(3)}_{2}=F^{(4)}_{2}=0$.  We
keep the term $\langle {\hat\vartheta}^{\dagger}(\r{1})
{\hat\vartheta}^{\dagger}(\r{2}) \rangle \langle {\hat\vartheta}(\rp{2})
{\hat\vartheta}(\rp{1}) \rangle$ from $F^{(4)}_{2}$ for retaining the
property of non-negative definition of the kernel (\ref{f2bog}).
\end{references}
\end{document}